# Human-AI Interaction: Evaluating LLM Reasoning on Digital Logic Circuit included Graph Problems, in terms of creativity in design and analysis.

Yogeswar Reddy Thota[1], Setareh Rafatirad[2], Homayoun Houman[2], Tooraj Nikoubin[1]

(1) Department of Electrical and Computer Engineering, University of Texas at Dallas, Richardson, USA
(2) Department of Electrical and Computer Engineering, University of California Davis, USA
YogeswarReddy.Thota@utdallas.edu, srafatirad@ucdavis.edu, hhomayoun@ucdavis.edu, Tooraj.Nikoubin@utdallas.edu

**ABSTRACT** Large Language Models (LLMs) are increasingly used by undergraduate students as on-demand tutors, yet their reliability on circuit- and diagram-based digital logic problems remains unclear. We present a human- AI study evaluating three widely used LLMs (GPT, Gemini, and Claude) on 10 undergraduate-level digital logic questions spanning non-standard counters, JK-based state transitions, timing diagrams, frequency division, and finite-state machines. Twenty-four students performed pairwise model comparisons, providing per-question judgments on (i) preferred model, (ii) perceived correctness, (iii) consistency, (iv) verbosity, and (v) confidence, along with global ratings of overall model quality, satisfaction across multiple dimensions (e.g., accuracy and clarity), and perceived mental effort required to verify answers. To benchmark technical validity, we applied an independent judge-based evaluation against official solutions for all ten questions, using strict correctness criteria. Results reveal a consistent gap between perceived helpfulness and formal correctness: for the most sequentially demanding problems (Q1-Q7), none of the evaluated LLMs matched the official answers, despite producing confident, well-structured explanations that students often rated favorably. Error analysis indicates that models frequently default to canonical textbook templates (e.g., standard ripple counters) and struggle to translate circuit structure into exact state evolution and timing behavior. These findings suggest that, without verification scaffolds, LLMs may be unreliable for core digital logic topics and can inadvertently reinforce misconceptions in undergraduate instruction.

## I. INTRODUCTION

Large Language Models (LLMs) have rapidly become part of the everyday learning workflow of undergraduate students. Tools such as ChatGPT, Gemini, and Claude are increasingly used as on-demand tutors for homework help, concept clarification, and exam preparation across computer science and engineering courses. Recent studies in computing education report that students frequently consult LLMs outside the classroom, often without formal guidance on their limitations or reliability [1- 3]. While early evidence suggests that LLMs can be helpful for programming syntax, conceptual explanations, and text-based problem solving [4], far less is known about their suitability for diagram-driven, sequential reasoning tasks, which are foundational in digital logic education.

This paper argues that contemporary LLMs systematically fail on undergraduate digital logic problems not due to missing domain knowledge, but due to a deeper reasoning mismatch: they substitute canonical textbook templates for schematic-grounded, constraint-driven reasoning. When faced with circuit diagrams, non-standard counters, or explicit feedback paths, models frequently default to familiar patterns (e.g., MOD-N counters or standard state machines) rather than deriving behavior from the actual circuit structure. This failure mode is consistent across models and persists even when explanations appear confident and well structured.

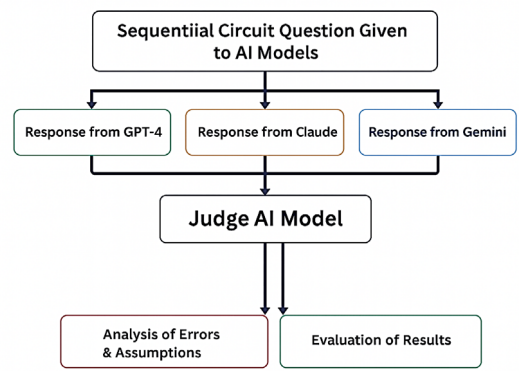

**Fig 1:** Human-AI evaluation workflow: student comparison of GPT, Claude, and Gemini responses followed by judge-based

Digital logic and digital systems courses occupy a unique position in the undergraduate curriculum. Unlike many introductory programming or mathematics problems, digital logic problems frequently rely on circuit diagrams, timing relationships, and explicit state evolution. Tasks such as analyzing counters, deriving state tables, completing timing diagrams, or determining frequency division behavior require students to reason over temporal sequences of states, propagate signal changes across clock edges, and respect exact structural constraints imposed by feedback paths and flip-flop configurations. Prior work in engineering education has shown that students themselves find topics such as finite-state machines (FSMs), asynchronous counters, and timing diagrams particularly challenging, even with traditional instruction [7- 8].

Existing evaluations of LLMs in education have largely focused on textual or algebraic problem domains, including programming exercises, mathematical word problems, and conceptual short-answer questions [4], [10- 12]. These studies often emphasize surface-level correctness, code compilation success, or student satisfaction. However, circuit-based reasoning introduces additional complexity: the model must correctly interpret diagrammatic structure, map that structure to formal sequential semantics, and maintain consistency across multiple time steps. Although recent work in multimodal learning has begun to explore how LLMs handle visual inputs [13- 15] there is limited empirical evidence examining how well such models perform on canonical undergraduate digital logic problems with well-defined official solutions, particularly when evaluated through both student perception and formal correctness.

Equally important is the role of human trust and confidence in LLM-generated answers. Prior human-centered AI research demonstrates that fluent, confident explanations can significantly increase user trust, even when the underlying content is incorrect [16- 18]. In educational settings, this raises a critical concern: students may accept confidently explained but incorrect answers without performing independent verification, especially for complex topics where manual simulation is cognitively demanding. Understanding this perception–correctness gap is essential for evaluating the educational risk of deploying LLMs in domains that require exact sequential reasoning.

In this work, we address these gaps through a human-AI evaluation study focused on undergraduate digital logic. We evaluate three widely used LLMs GPT, Gemini, and Claude on 10 undergraduate-level digital logic questions covering non-standard counters, JK-based state transitions, timing diagrams, frequency division, and FSM derivation. Twenty-four undergraduate students participated in a pairwise comparison study, providing per-question feedback on model preference, perceived correctness, consistency, verbosity, and confidence. In addition, students reported global satisfaction metrics and the mental effort required to understand or verify model answers. To complement student perception, we apply an independent judge-based evaluation against official solutions, enabling a direct comparison between student trust and formal correctness.

This study is guided by the following research questions:
(1) *Can undergraduate students reliably depend on LLMs to solve circuit- and state-machine-based digital logic problems?*
(2) *Do student preferences and confidence judgments align with formal correctness as defined by official solutions?*
(3) *What types of digital logic problems most consistently expose LLM failure modes?*
(4) *How does explanation confidence influence student trust and perceived cognitive effort?*

The contributions of this paper are threefold. First, we present a human-centered evaluation of LLMs on circuit-based undergraduate digital logic problems, a domain that has received limited attention in prior LLM-in-education research. Second, by incorporating judge-verified correctness, we expose systematic discrepancies between student perception and actual model reliability. Third, we identify recurring failure patterns in sequential reasoning, including default reliance on canonical textbook structures and difficulty translating circuit diagrams into exact temporal behavior. Together, these findings provide actionable insights for educators, students, and tool designers regarding the responsible use of LLMs in digital logic education.

Together, these considerations motivate a systematic investigation of how and why LLMs fail on circuit- and state-machine-based digital logic problems, despite appearing confident and persuasive. To situate this study within existing work, the following section reviews prior research on LLM use in education, student reasoning in digital logic, and human-centered evaluations of AI explanations, highlighting where current evidence remains insufficient for diagram-grounded reasoning tasks.

## II. Literature Review

### A. *LLMs in Undergraduate Education*

The adoption of LLMs in undergraduate education has accelerated rapidly, particularly following the public release of conversational AI systems. Recent studies report widespread, informal use of LLMs by students for homework assistance, concept clarification, and exam preparation across computing disciplines [1- 3]. While early findings suggest that students perceive these tools as helpful and time-saving, concerns remain regarding over-reliance, academic integrity, and the accuracy of generated explanations.

Within computing education, prior work has primarily examined LLM performance on programming-related tasks, including code generation, debugging, and short-answer conceptual explanations [4- 6]. These studies generally report moderate success on syntactic and well-scoped problems, but also document frequent hallucinations and subtle logical errors. Importantly, much of this literature emphasizes student satisfaction and usability, rather than rigorous verification against authoritative solutions.

### B. *Digital Logic as a Distinct Educational Domain*

Digital logic and digital systems courses differ fundamentally from many other undergraduate computing subjects. Canonical textbooks emphasize the importance of precise state transitions, clock-driven behavior, and strict adherence to circuit structure [7,8]. Prior education research has shown that students struggle disproportionately with topics such as finite-state machines, asynchronous counters, and timing diagrams, where errors often stem from incorrect mental models of state evolution [9].

In many digital logic contexts, solutions are evaluated against exact state transitions or timing behavior, where even small deviations can invalidate the intended circuit operation. As a result, pedagogical tools that provide plausible but incorrect reasoning may be especially harmful. Despite this, there is limited empirical evidence evaluating whether modern LLMs can meet the strict correctness demands imposed by undergraduate digital logic problems.

### C. *Human-Centered Evaluation of LLM Outputs*

Beyond raw accuracy, recent AI research emphasizes the importance of human-centered evaluation, recognizing that user trust and perception often diverge from formal correctness [10- 12]. Studies consistently show that fluent, confident explanations increase perceived reliability, even when answers are incorrect. This divergence between perceived reliability and formal correctness motivates the need for evaluations that explicitly contrast user judgment with ground-truth performance.

Human evaluations commonly employ pairwise comparisons, Likert-scale judgments, and preference-based metrics to capture these effects [11]. However, few studies combine such evaluations with ground-truth verification, especially in technically exact domains such as digital logic.

### D. *Multimodal and Diagram-Based Reasoning in LLMs*

Circuit-based problems introduce additional challenges due to their reliance on visual structure and diagrammatic reasoning. While recent multimodal models demonstrate progress in image- text understanding, multiple studies report persistent limitations when reasoning over structured diagrams and symbolic representations [13- 15]. These limitations often manifest not as misrecognition of components, but as failures to correctly translate visual structure into formal reasoning steps.

Such findings suggest that even when LLMs can identify circuit elements and describe their function, they may still fail to produce correct sequential behavior an issue directly relevant to digital logic education.

### E. *Trust, Confidence, and Educational Risk*

Prior work on human-AI interaction demonstrates that explanation fluency and expressed confidence can influence user trust independently of correctness [16–18]. In educational settings, this raises concerns about how students calibrate trust when evaluating complex, constraint-driven answers. In educational settings, this can lead to the reinforcement of misconceptions, particularly for topics requiring precise reasoning.

These findings motivate the need for systematic evaluation of LLMs that jointly considers correctness, confidence, and student perception, especially in domains where formal accuracy is essential. Collectively, prior work demonstrates both the rapid adoption of LLMs in education and persistent challenges in student reasoning for state-based and diagram-driven topics. At the same time, human-centered studies reveal that confidence, explanation fluency, and trust often diverge from formal correctness. However, these strands of research remain largely disconnected: few studies jointly examine student perception, formal correctness, and circuit-specific reasoning within a single evaluation framework. This gap directly motivates the human- AI methodology described in the next section, which is designed to capture both perceived usefulness and verified correctness for undergraduate digital logic problems.

## III. Methodology

This study employs a human-AI evaluation framework to assess how reliably large language models perform schematic-grounded, constraint-driven reasoning on undergraduate-level digital logic problems. The methodology combines pairwise student evaluations, multi-dimensional feedback instruments, and an independent judge-based correctness assessment against official solutions.

### A. Participants

Twenty-four undergraduate students participated in the study. All participants were enrolled in, or had recently completed, a course covering digital logic and digital systems, including topics such as counters, flip-flops, timing diagrams, and finite-state machines. Participation was voluntary, and responses were collected anonymously. Students were not incentivized based on performance and were instructed that the study evaluated model behavior, not their own correctness.

### B. Question Set

The evaluation used ten undergraduate-level digital logic questions (Q1- Q10). All questions were drawn from standard instructional material and reflect common assessment items in digital systems courses. The questions required analysis of circuit diagrams and state-based behavior, including:

- Non-standard asynchronous counters with feedback
- JK flip-flop- based state tables
- Timing diagram completion
- Non-binary frequency division
- FSM derivation from state diagrams

Each question had a single official solution, provided in course materials, which served as the ground truth for judge-based evaluation. For the purposes of this study, a model response is considered a *failure* if it deviates in any way from the official solution provided in course materials, including incorrect state sequences, missing or extra transitions, incorrect timing behavior, or incorrect frequency ratios. No partial correctness was awarded, as even a single incorrect transition renders a digital logic solution invalid for instructional or assessment purposes.

### C. Large Language Models Evaluated

We conducted all model evaluations using EduArena, a unified evaluation platform that enables controlled comparison of multiple LLMs. Three widely used LLMs Claude (Haiku), ChatGPT (GPT-5.2 Pro), and Gemini (version 3 Pro) were evaluated with 24 students who took Digital Circuit, undergraduate level course, and the responses of these models evaluated with Perplexity as a judge model:

Because the study was designed to reflect realistic student behaviour, participants were aware of the model identities they were evaluating (e.g., GPT, Gemini, Claude). Students explicitly selected which pair of models they wished to compare, based on familiarity, prior experience, or personal preference. This design intentionally captures the role of brand recognition and expectation effects in shaping trust, preference, and perceived correctness.

To reduce cognitive overload and direct multi-way comparison bias, students did not evaluate all three models simultaneously. Instead, they performed pairwise comparisons (e.g., GPT vs. Gemini or Gemini vs. Claude) for each question. Within each comparison, responses were labeled as *Model A* and *Model B* only for response organization; students were fully aware of which underlying AI system corresponded to each label. The mapping between labels and model identities was preserved during analysis.

This approach balances ecological validity mirroring how students naturally choose and trust AI tools with structured comparison, enabling analysis of how brand familiarity can influence confidence, preference, and perceived correctness even when formal correctness is absent.

### D. Pairwise Evaluation Design

For each question, students compared the outputs of two selected models and provided feedback across five dimensions:

1. **Preferred Model** - Which model's response they preferred overall
2. **Perceived Correctness** - Which model they believed produced the correct answer
3. **Consistency** - Whether the models' reasoning and conclusions appeared consistent
4. **Verbosity** - Whether any response contained unnecessary or overly verbose content
5. **Confidence** - Whether the models appeared confident in their answers

Each feedback item allowed four options: Model A, Model B, Both, or None of them. This design captured nuanced perceptions, including cases where students judged both models as equally good or equally poor.

### E. Evaluation Metrics

In addition to per-question feedback, students provided global assessments after completing the ten questions:

- **Overall Best Model** (single choice)
- **Overall Efficiency Rating** - How accurate, clear, and to-the-point the preferred model was

- **Satisfaction Ratings** (Likert scale) across multiple dimensions, including accuracy, consistency, clarity, structure, depth, and helpfulness
- **Mental Effort** - A Net Promoter Score- style measure indicating how much effort was required to understand or verify the model's answers

These measures captured cumulative trust, usability, and cognitive load, complementing per-question evaluations.

*F.    Judge-Based Correctness Evaluation*

To assess formal correctness, an independent judge evaluation was performed for all ten questions (Q1- Q10). Model outputs were compared directly against the official solutions provided in the course materials. Correctness was evaluated strictly, with no partial credit awarded. A response was marked correct only if it fully matched the official answer in terms of sequence, state transitions, timing behavior, or frequency ratio. This judge-based assessment served as a ground-truth reference, enabling direct comparison between student perception and formal correctness.

*G.    Data Aggregation and Normalization*

All responses were aggregated per question and per feedback dimension. For Questions 1- 5, each feedback item contained 24 student responses, including selections of "All" and "None of them," which were treated as valid outcomes. Percentages were normalized accordingly. Model A and Model B labels were mapped back to actual model identities prior to analysis.

*H.    Analysis Approach*

The analysis focused on three primary comparisons:

1. **Student Perception vs. Judge Correctness** - Identifying alignment and mismatch
2. **Confidence- Correctness Relationship** - Examining cases of confident but incorrect responses
3. **Problem-Type Sensitivity** - Comparing model behavior across different categories of digital logic problems

Descriptive statistics were used throughout, with emphasis on patterns and failure modes rather than inferential claims, given the exploratory nature of the study. This methodology enables a structured comparison between student judgments and formal correctness by combining pairwise evaluations, multi-dimensional feedback, and judge-based verification against official solutions. By explicitly accounting for preference, confidence, and rejection alongside strict correctness criteria, the design captures both how models are perceived and how they actually perform on digital logic tasks. The following section presents the resulting empirical findings, highlighting where student trust aligns or diverges from verified model reliability.

## IV. RESULTS AND ANALYSIS

This section presents the results of the human- AI evaluation across the first ten undergraduate digital logic questions (Q1- Q10). We first analyze overall student perception using absolute aggregation, then contrast these perceptions with judge-verified correctness for Q1- Q10. Finally, we examine the role of confidence, ambiguity, and rejection in student trust. Across these analyses, we observe a recurring pattern in which student trust and perceived correctness diverge sharply from formal correctness, particularly for diagram-intensive sequential reasoning tasks.

*A.    Overall Student Preference and Acceptance*

Figure 2 e) presents the absolute aggregation of student preferences across all ten questions. GPT receives the highest number of preference credits, followed by Claude and Gemini. However, two additional signals are critical. First, the **"All models"** category has a non-trivial count, indicating that students frequently perceived *both* compared models as equally acceptable. Second, a substantial number of responses fall under **"None of them"**, showing that students explicitly rejected all provided answers for several questions. This pattern suggests that while GPT is most often preferred, student trust is not absolute. Preference frequently coexists with uncertainty or dissatisfaction, especially for questions involving non-standard counters and state-machine reasoning.

*B.    Perceived Correctness vs. Formal Correctness*

Figure 2 d) summarizes student judgments regarding which model produced the correct answer. GPT again leads in perceived correctness, with Claude and Gemini closely following. However, when compared against the judge-verified evaluation for, a stark discrepancy emerges: none of the evaluated models produced a fully correct solution for most of the questions as shown in Figure 3 and Gemini is doing better by giving correct response compared to other models in this experiment when compared with the ground truth. Across Q1–Q7, 68% of student judgments labeled an incorrect response as correct, despite zero models matching the official solutions.

Despite this, students consistently selected models most often GPT as producing the correct answer. This mismatch demonstrates that perceived correctness is driven by explanation plausibility rather than formal alignment with circuit behavior. Even when answers deviated significantly from the official state sequences or timing diagrams, students frequently labeled them as correct.

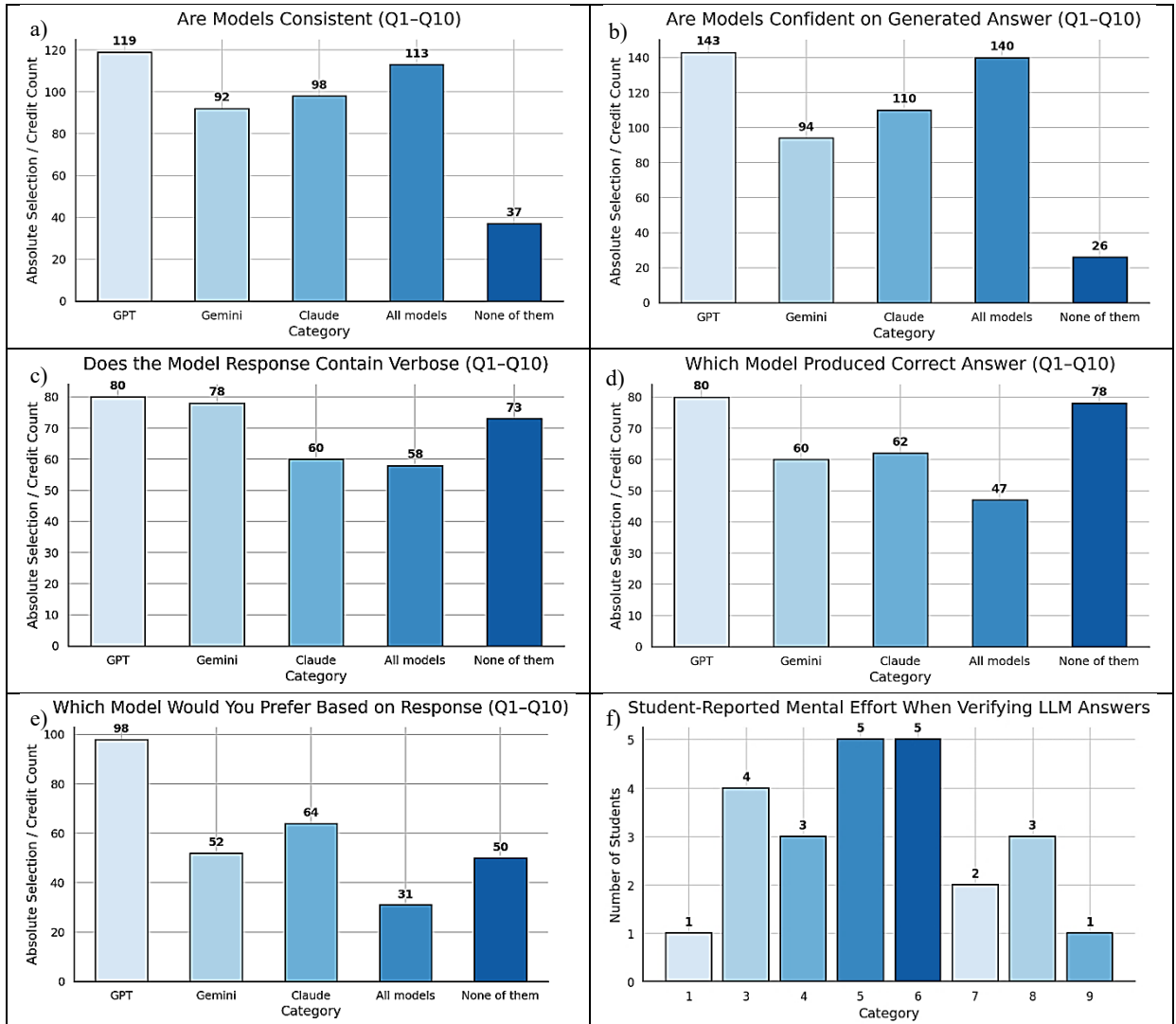

Fig 2: Aggregated student feedback across all ten questions based on AI models responses in this study

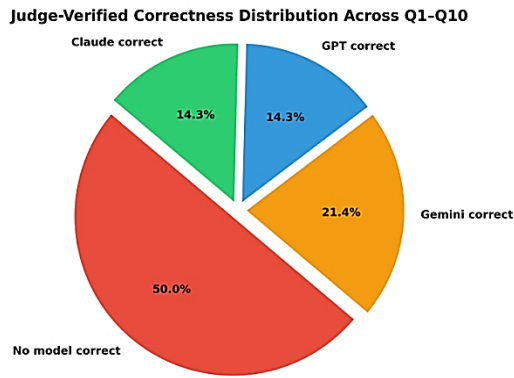

Fig 3: Judge Model feedback on model responses considering ground truth answers.

### C. Consistency and Explanation Structure

Figure 2 a) reports student judgments of model consistency. GPT is rated as the most consistent model, followed by Claude and Gemini. Consistency here reflects perceived internal coherence and narrative flow, not logical correctness. Many incorrect responses still received high consistency scores, reinforcing the observation that students equate smooth explanation with reliable reasoning. This finding is particularly concerning for digital logic education, where incorrect assumptions about clocking, feedback, or state transitions can propagate subtle but critical misunderstandings.

### D. Verbosity and Confidence Effects

Figure 2 c) shows the distribution of responses identifying unnecessary or overly verbose content. GPT and Gemini are most frequently marked as verbose, while Claude is perceived as more concise. However, verbosity alone does not reduce trust; in many cases, verbose explanations coincided with higher preference and confidence scores.

Figure 2 b) presents perceived model confidence, where GPT again dominates. Importantly, confidence scores remain high even when students selected "None of them" for correctness or preference on the same questions. This indicates that confidence and correctness are largely decoupled in student evaluation.

Figure 2 e) is the reported level of stress students felt while evaluating the LLM model responses that reveals how difficulty students felt during this evaluation. Together, these figures reveal a critical phenomenon: confident, well-structured explanations can mask deep logical errors, especially in diagram-driven problems where students may not independently simulate behavior.

### E. Role of "All Models" and "None of Them"

Across all feedback dimensions, both All models and None of them appear with meaningful frequency. "All models" selections indicate ambiguity students accept multiple answers without strong discrimination while "None of them" captures explicit recognition of failure. The coexistence of these categories highlights that students often sense *something is wrong* without being able to identify which model, if any, is correct. This ambiguity is most pronounced in questions requiring multi-step temporal reasoning, such as asynchronous counters and FSM timing.

Overall, the results reveal a systematic disconnect between student perception and formal correctness, particularly for problems requiring non-standard sequential reasoning and precise state evolution. While GPT is most frequently preferred and rated as confident and consistent, none of the evaluated models produced correct solutions for the most diagram-intensive questions. These findings raise important questions about *why* LLMs fail in such settings and *which aspects of digital logic reasoning are most problematic*. The next section addresses these questions through a detailed error analysis of model behavior across the evaluated problem types.

## V. ERROR ANALYSIS: WHY THE MODELS FAILED (AND WHEN THEY SUCCEEDED)

The aggregate results in Section 4 establish a clear gap between student perception and judge-verified correctness, but they do not by themselves explain *why* LLMs fail on certain digital logic questions and succeed on others. In this section, we analyze the ten evaluated questions (Q1- Q10) to identify recurring failure modes and relate them to specific structural properties of the problems.

### A. Non-Standard Sequential Behavior (Q1, Q3)

Questions 1 and 3 required analysis of non-standard sequential behavior defined by specific circuit wiring and timing constraints. In Question 1, the official solution exhibited a non-binary counting sequence 0→6→2→5→1→4→0→0, driven by asynchronous feedback paths in a three-flip-flop JK counter. Correct analysis requires (i) reading the J, K, and clock connections for each flip-flop, (ii) constructing the corresponding state-transition table, and (iii) enumerating the reachable states starting from 000. Figure 4 represents the circuit diagram that was included in question 1.

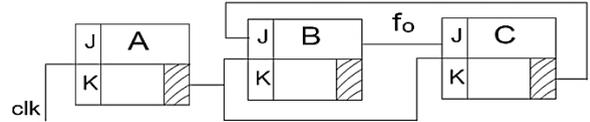

Fig 4: Binary counter circuit included in question 1 considering main outputs are ABC from flip flops, and required to find what is the count routine of this binary counter

None of the models performed this circuit-grounded reasoning. GPT and Claude both declared the circuit to be a "3-bit binary up-counter" with all JK inputs effectively tied high and produced the standard MOD-8 sequence 0- 7. Gemini went slightly further by arguing that additional gating truncated the count at 5, yielding a MOD-6 sequence 0- 5, but it still assumed a conventional monotone up-counter. In all three cases, the models overwrote the non-standard feedback with a canonical ripple-counter template rather than deriving next states from the actual schematic. Follow-up conversation has been made with the models to make them understand the circuit which still failed to get the exact circuit diagram, this is explained in detail in Table 1.

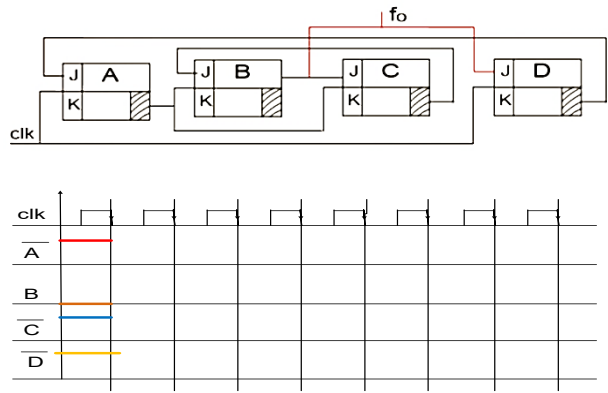

Fig 5: Flip-flop sequential circuit and timing diagram included in question : which has four JK-MS-FFs, if K(A)=K(B)=K(C)=K(D)= 1 and asked to complete the timing diagram for (A', B, C' & D') output signals. Initial values for all flop flops are reset.

A similar failure mode appears in Question 3, which required students to complete timing diagrams for a four-flip-flop sequential circuit with explicit non-standard feedback. The correct solution depends on propagating signal transitions through the specific red feedback connection shown in Figure 6, resulting in a custom, non-binary timing sequence. In contrast, all three evaluated models interpreted the circuit as a conventional 4-bit MOD-16 counter or simplified it into a short repeating loop, effectively ignoring the circuit's actual feedback wiring. Figure 2 shows that students frequently judged these answers as correct or preferable despite their complete mismatch with the official sequences, underscoring how fluent explanations can mask structural errors.

### B. State-Transition Grounding Failures (Q2, Q4, Q5)

Questions 2, 4, and 5 required derivation of exact state tables or transitions from diagrams or excitation constraints. For Q2, the official answer specifies next-state mappings and output values for all combinations of ABC and input X under fixed JK conditions. The three models each produced a full 16-row table, and the tables were internally consistent, but none matched the official transitions. Instead of simulating the actual circuit, the models inferred plausible JK behavior (e.g., "flip-flops toggle when J=K=1") and filled in state transitions to fit that intuition.

Similar issues appear in Q4 and Q5. In Q4, which extends the Q3 circuit and asks for both timing and frequency behavior, the models again imposed simple counter semantics rather than following the given feedback through multiple clock edges. In Q5, where the task is to derive a state table from a state diagram as shown in Figure 6, the models sometimes produced syntactically valid but misaligned tables, suggesting that they recognized the diagrammatic form but did not consistently track state labels and transitions across the diagram.

Across these questions, the models are capable of generating *some* state table that makes sense locally, but they struggle to ground every entry in the exact diagram or excitation constraints, leading to systematic deviations from the official answers.

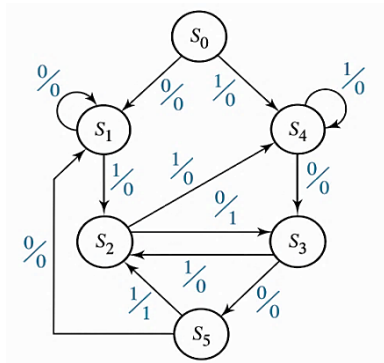

Fig 6: Given state diagram in question 5 to determine the state table [19].

### C. Fixed-Constraint Synthesis Failures (Q6, Q7)

Questions 6 and 7 required designing frequency dividers using three JK master-slave flip-flops under strict excitation constraints. In Q6, the official design implements a divide-by-5 circuit with K(A)=K(B)=K(C)=J(B)=1; Q7 specifies a divide-by-6 circuit with K(A)=K(B)=K(C)=J(A)=1. These constraints effectively force particular internal connections, and the grading rubric treated any deviation as incorrect.

All three models successfully synthesized divide-by-5 and divide-by-6 counters at a functional level. Claude and Gemini proposed synchronous MOD-5 and MOD-6 counters with derived J/K equations, while GPT suggested ripple counters with reset logic at specific states. However, none of these designs respected the fixed *K* and *J* constraints given in the problem statements. That is, the models optimized solely for *functional equivalence* ("implement any divide-by-N counter"), not for *structural equivalence* to the required architecture.

This behavior illustrates a different failure mode: when a problem demands adherence to a particular implementation template, LLMs tend to produce a valid design from first principles rather than honoring the stated structural constraints.

### D. Canonical Pattern Recognition Success (Q8-Q10)

Model performance improved markedly on Questions 8-10, which correspond closely to canonical patterns in digital logic textbooks. Question 8 involved a two-flip-flop circuit where the correct answer is a frequency doubling behavior $F_{OUT}/F_{IN}=2$. Only Gemini identified this dual-edge detection pattern and derived the correct ratio; GPT and Claude focused on synchronizer and divide-by-four interpretations and centered their answers around $F_{OUT}/F_{IN}=1$ or ¼. Questions 9 and 10 were more strongly aligned with well-known templates. Q9 asked for a low-to-high edge detector using two D flip-flops and a fast clock; Q10 extended this to detect both rising and falling edges. The official solutions use a two-stage synchronizer with output

$$Y=Q1 \cdot \bar{Q}2 \text{ (Q9) and } Y=Q1 \oplus Q2 \text{ (Q10)}.$$

All three models reproduced these circuits exactly, connecting D1=Xin, D2=Q1, clocking both flip-flops with the fast clock, and deriving the same Boolean expressions as in the official diagrams. Students accordingly rated these answers as highly correct and helpful (Figures 3 and 5), and judge-based evaluation confirmed perfect agreement with the ground truth. These results indicate that when a problem can be mapped directly to a memorized architectural pattern such as a two-FF edge detector current LLMs can be highly reliable.

Table 1: Comparative failure analysis of GPT-4, Claude, and Gemini on a diagram-based sequential logic problem on question 1. While all models exhibit strong local reasoning, each introduces implicit timing or structural assumptions not justified by the diagram, leading to divergent and incorrect state sequences.

| Dimension | GPT-4 (OpenAI) | Claude (Anthropic) | Gemini (Google) |
|---|---|---|---|
| Initial interpretation | Assumed a standard 3-bit ripple MOD-8 up-counter | Assumed a standard MOD-8 binary counter | Assumed a ripple down-counter based on $\bar{Q}$ clocking |
| Primary incorrect assumption | *Implicitly assumed J = K = 1 for all flip-flops* | *Implicitly assumed tied J/K inputs imply toggle behavior* | *Assumed textbook ripple timing semantics without validating sampling order* |
| Netlist reconstruction behavior | Initially hallucinated J/K connections; later admitted UNKNOWN when constrained | Confidently asserted J/K values without diagram evidence | More cautious, but still inferred J/K ties and clock polarity |
| Clocking model assumed | Mixed synchronous and asynchronous updates inconsistently | Fully asynchronous ripple, but with parallel evaluation | Fully asynchronous ripple with sequential propagation |
| Critical timing assumption | Treated A and C as updating simultaneously on clk↓ | Treated all FF updates as textbook ripple with immediate propagation | Treated C as sampling *post-update* values of A and B |
| Where reasoning breaks | Violates causal ordering: uses updated signals to justify prior clock events | Collapses underspecified wiring into canonical counter patterns | Correctly simulates ripple, but assumes visibility of updated upstream states |
| Response when constrained by strict netlist rules | Eventually acknowledged hidden assumption and revised reasoning | Continued to rely on structural analogies (binary counter) | Explicitly stated that the official sequence requires a different timing model |
| Ability to derive official sequence (0→6→2→5→1→4→0) | Only after being told the answer and analyzing assumption failure | Could not reconcile sequence with its assumed counter model | Explicitly stated that the diagram alone cannot determine the needed timing |
| Self-diagnosed failure mode | Incorrect update ordering (simultaneous vs ripple) | Over-reliance on canonical counter archetypes | Diagram underspecifies timing; multiple valid interpretations exist |
| Core limitation exposed | Hallucinates logical structure under ambiguity | Overgeneralizes from known circuit templates | Requires timing semantics not present in the diagram |
| Key insight for this study | Logical correctness ≠ semantic correctness | Structural familiarity ≠ diagram understanding | Correct simulation still fails without intent inference |

### E. Summary of Failure Modes and Success Conditions

Across all ten questions, model failures are not random. Two overarching patterns emerge:

**Template substitution over circuit grounding.** For non-standard counters (Q1, Q3), JK-based state tables (Q2, Q4), and constrained dividers (Q6, Q7), the models tended to recognize a high-level "type" of circuit (binary counter, JK counter, mod-N divider) and generate behavior consistent with that type, even when it conflicted with the actual schematic or constraints. This often produced fluent, internally consistent explanations that students perceived as correct, despite substantial deviations from official solutions.

**Canonical pattern success**. When the target circuit coincided with a widely taught template especially the two-FF edge-detector patterns in Q9 and Q10 all three models produced fully correct designs. In these cases, the same pattern-matching behavior that caused hallucinations in Q1- Q7 became an asset.

Taken together, these findings support the central claim of this paper: current LLMs act more as persuasive explainers than reliable solvers for circuit-specific digital logic problems. They are most effective when instantiating canonical patterns and least reliable when correctness depends on faithfully simulating an idiosyncratic schematic or satisfying fixed excitation constraints.

### F. Judge Model Observations:

Below are the observations that are stated by judge model based on the analysis:

<u>Perplexity</u>: *Judge-based evaluation across all ten questions reveals stark performance disparities: none of the three models (GPT, Gemini, Claude) produced fully correct solutions for Q1- Q7, which demanded strict adherence to non-standard circuit wiring, JK excitation constraints, or explicit timing behavior—these structurally constrained problems yielded a 0% success rate (0/21 model-question pairs). In contrast, Q8 saw partial success with only Gemini correctly identifying the dual-edge frequency-doubling pattern ($F_{OUT}/F_{IN}=2F_{OUT}/F_{IN}=2$), while Q9- Q10 elicited perfect performance from all models, reproducing the canonical two-D-FF edge detectors exactly (6/6 correct). Overall accuracy across 30 model-question attempts was 23% (7/30 correct), with questions exhibiting at least one correct model limited to 30% (3/10). GPT emerged as the student-preferred model due to its consistent fluency and structure, though it was frequently rated most verbose; Claude balanced conciseness with*

*reliability on canonical patterns; Gemini excelled technically on Q8 but trailed in student perception. These patterns underscore that while GPT maximizes perceived helpfulness, no model can be trusted without verification for diagram-intensive sequential reasoning.*

## VI. CONCLUSION

This study examined whether undergraduate students can reliably depend on LLMs for solving digital logic problems involving combinational circuits, sequential circuits, counters, and state machines. Using a controlled set of ten undergraduate-level questions, we combined large-scale student feedback with a detailed judge-verified comparison against official textbook and instructor solutions. Our findings reveal a consistent and important gap between perceived usefulness and actual correctness of LLM-generated answers.

Across the first seven questions, which required strict adherence to given circuit diagrams, fixed flip-flop excitation constraints, or explicit timing behavior, none of the evaluated models (GPT, Gemini, or Claude) reproduced the official solutions. Instead, the models consistently generated functionally plausible but structurally different designs, often replacing the given circuit with a generic counter or state machine that satisfied the high-level goal but violated specific wiring or excitation constraints. This behavior occurred despite high confidence and detailed explanations in the model outputs, demonstrating that fluency and apparent correctness are insufficient indicators of reliability for circuit-specific reasoning tasks.

In contrast, model performance improved markedly for Questions 9 and 10, which corresponded to canonical edge-detection patterns using two D flip-flops. For these problems, all models produced solutions that matched the official designs exactly. Question 8 represented an intermediate case, where only one model, Gemini correctly identified the dual-edge detection configuration required to achieve the specified frequency ratio. These results indicate that current LLMs are most reliable when the problem aligns with well-known, frequently repeated circuit patterns, and substantially less reliable when reasoning must be grounded in a particular schematic or non-standard sequential behavior.

Student feedback further highlights the educational risk of this mismatch. Across all ten questions, students frequently rated model responses as confident, clear, and preferable, even in cases where none of the models were correct according to the official solutions. The frequent selection of "Both" or "All models" in preference questions, combined with the high "None of them" rates in correctness judgments, suggests that students struggle to independently verify LLM outputs for sequential logic problems. This reinforces the concern that LLMs may inadvertently amplify misconceptions in digital logic education when used without structured verification.

Taken together, these findings suggest that current LLMs should not be relied upon as authoritative solvers for circuit-specific digital logic problems, particularly those involving fixed excitation constraints, non-binary counting sequences, or explicit timing analysis. However, LLMs can still play a valuable role as supplementary tools for explaining concepts, reviewing canonical patterns, or providing alternative perspectives provided their outputs are carefully validated against schematics and formal analysis methods.

Future work may explore whether tighter multimodal grounding, explicit constraint enforcement, or stepwise simulation prompts can improve model reliability for such tasks. Until then, instructors and students should treat LLM-generated circuit solutions as advisory rather than definitive, especially in courses where precision and structural correctness are essential.

## ACKNOWLEDGMENT

The authors would like to acknowledge Turing, for providing access to EduArena (https://www.eduarena.ai/), the evaluation platform used in this study. EduArena supports multiple chat modes and enables side-by-side interaction with leading LLMs, including ChatGPT (OpenAI), Claude (Anthropic), and Gemini (Google). We specifically utilized the side-by-side chat mode, which allowed students to compare responses from different models simultaneously for the same question, facilitating direct and informed evaluations of model behavior. This interface played a critical role in enabling fair, controlled, and user-friendly human- AI comparisons. The authors also thank Turing, Inc. for sponsoring this research and supporting the exploration of LLMs in educational contexts. The database of questions considered and responses of students has been provided here: https://github.com/SNS-LAB-AI/LLM_Evaluation that can be used for further research works.

**Competing Interests:** On behalf of all authors, the corresponding author states that there is no conflict of interest.
- Funding Information (NA)
- Author contribution (NA)
- Data Availability Statement (NA)
- Research Involving Human and /or Animals (NA)
- Informed Consent (NA)